\documentclass[twocolumn,showpacs,preprintnumbers,amsmath,amssymb]{revtex4}

\usepackage{graphicx}
\usepackage{graphics}
\usepackage{dcolumn}
\usepackage{bm}

\begin{document}

\preprint{APS/123-QED}

\title{Ionization of hydrogen atoms by electron impact
at 1eV, 0.5eV and 0.3eV above threshold}

\author{J. N. Das  and S. Paul}
\affiliation{Department of Applied Mathematics, University College
of Science, 92 A. P. C. Road, Calcutta 700 006, India.}
\author{K. Chakrabarti}
\email{kkch@eth.net} \affiliation{Department of Mathematics,
Scottish Church College, 1 \& 3 Urquhart Square, Calcutta 700 009,
India}

\date{\today}

\begin{abstract}
We present here triple differential cross sections for ionization
of hydrogen atoms by electron impact at 1eV, 0.5eV and 0.3eV
energy above threshold, calculated in the hyperspherical partial
wave theory. The results are in very good agreement with the
available semiclassical results of Deb and Crothers \cite{DC02}
for these energies. With this, we are able to demonstrate that the
hyperspherical partial wave theory yields good cross sections from
30 eV \cite{DPC03} down to near threshold for equal energy sharing
kinematics.
\end{abstract}


\maketitle

\section{Introduction}
    Over the last couple of years considerable progress has been
made in the study of ionization of hydrogen atoms by electron
impact, apparently the simplest three body coulomb problem in
quantum scattering theory, at low energies. Still full
understanding of this problem, particularly near threshold, has
not been achieved.

Ionization near threshold was studied by Pan and Starace
\cite{PC92} where they reported a distorted wave calculations of
(e, 2e) process with H, He and other rare gas targets at excess
energies 4 eV (above the respective ionization thresholds) and
below for equal energy sharing kinematics. For atomic hydrogen
they reported results at excess energies 2 eV and 0.5 eV above
threshold in the coplanar constant $\theta_{ab}$ geometry (in
which the angle $\theta_{ab}$ between the emerging electrons
remain fixed) with $\theta_{ab}=\pi$. Jones, Madison and
Srivastava \cite{JMS92} also reported a distorted wave \linebreak
(e, 2e) calculation with atomic hydrogen and helium target for
equal energy sharing kinematics but different geometries. Their
results were in good qualitative agreement with the experiments at
2 eV above threshold for atomic hydrogen for constant
$\theta_{ab}$ geometries. However, for the other geometries
presented, there were considerable deviations from the
experimental results.

For equal-energy-sharing kinematics McCurdy and co-workers made a
break-through calculation in their exterior complex scaling (ECS)
approach \cite{RB99,BR01a,BR01b}. Their results for 30eV, 25eV,
19.6eV and 17.6eV agree excellently with the measured results of
R\"oder \textit{et al} \cite{RE97,RB02,RR96}. However, below 2eV,
results of ECS theory are not yet available. Another sophisticated
approach is the convergent close-coupling (CCC) theory of Bray and
associates \cite{BK94,IB99,IB00}, which works beautifully for many
atomic scattering problems and reproduces ionization cross section
results very satisfactorily above 2eV excess energy. For 2eV
excess energy their results differ approximately by a factor of
two from the absolute measured values \cite{RE97}. Below 2eV
excess energy CCC results are also not available. Recent
calculations of Das and co-workers \cite{DPC03} for equal energy
sharing kinematics in the hyperspherical partial wave (HPW) theory
also reproduced the experimental data \cite{RE97,RR96} quite
satisfactorily.

So far the hyperspherical partial wave theory has not been tested
below 2eV excess energy. Deb and Crothers \cite{DC02} have
reported a semiclassical calculation that gives very good cross
section results for low energies of 4eV and 2eV above threshold
and also for  energies 1eV, 0.5eV and 0.3eV above threshold. This
calculation encouraged us to test whether the hyperspherical
partial wave theory works for excess energy below 2eV. Here we
made such a calculation for excess energy of 1eV, 0.5eV and 0.3eV
above the ionization threshold. We found that the HPW theory gives
cross section results in very good agreement with the semi
classical calculation of Deb and Crothers \cite{DC02} for the
above energies. One only needs to increase the asymptotic range
parameter $R_\infty$ (defined later) to sufficiently large values
of several thousands of a.u. It is interesting to note here that
the hyperspherical $\mathcal{R}$-matrix with semiclassical
outgoing waves(H$\mathcal{R}$M-SOW) calculation of Selles
\textit{et al} \cite{SM02}, for the double photo-ionization of the
Helium atom, also requires  $R_\infty$ values of several thousands
of a.u. for converged results.

\section{Hyperspherical Partial Wave Theory}
   The hyperspherical partial wave theory has been described in details
in \cite{JND98, DPC03} and briefly in \cite{JND01,JND02}. In this
approach we determine scattering amplitude from the T-matrix
element given by
\begin{equation}
T_{fi} = \langle \Psi_f^{(-)}|V_i|\Phi_i \rangle
\end{equation}
where $\Phi_i$ is the initial state wave function, $V_i$ is the
interaction potential in this channel and $\Psi_f^{(-)}$ is the
exact two-electron continuum wave function with incoming boundary
conditions in presence of the  nucleus, which is considered
infinitely heavy and stationary at the origin. The scattering
state $\Psi_f^{(-)}$ is determined by expanding it in terms of
symmetrized hyperspherical harmonics \cite{JND98,LIN74} which are
functions of five angular variables. The hyperradius and the
angular variables are defined by $R=\sqrt{{r_1}^2 + {r_2}^2}$,
$\alpha = \arctan(r_2/r_1), \; \hat{r}_1=(\theta_{1}, \phi_{1}),
\; \hat{r}_2=(\theta_{2}, \phi_{2})$ and $\omega=(\alpha,
\hat{r}_1, \hat{r}_2)$ and set $P=\sqrt{{p_1}^2 + {p_2}^2}$,
$\alpha_0 = \arctan(p_2/p_1), \; \hat{p}_1=(\theta_{p_1},
\phi_{p_1}), \; \hat{p}_2= (\theta_{p_2}, \phi_{p_2})$ and
$\omega_0=(\alpha_0, \hat{p}_1, \hat{p}_2)$, $\vec{p}_1, \;
\vec{p}_2$ being momenta of the two outgoing electrons of energies
$E_1$ and $E_2$, and positions $\vec{r}_1$ and $\vec{r}_2$.

Different sets of radial waves with definite $\mu = (L, S, \pi)$,
(where L is the total angular momentum, S is total spin and $\pi$
is the parity) satisfy different sets of coupled equations each of
the form
\begin{equation}
\Big[ \frac{d^2}{dR^2} + P^2 - \frac{\nu_N\,(\nu_N+1)}
{R^2}\Big]f_N + \sum_{N'} \frac{2P\;\alpha_{NN'}}{R} \, f_{N'} =
0.
\end{equation}

These equations are truncated to N = $N_{mx}$ which is the number
of channels retained in the calculation for each fixed $\mu$. The
$N_{mx}$ number of independent solutions of the truncated
equations, need to be determined over the interval $(0, \infty)$.
These equations may then be solved in different alternative
approaches. One possibility is to partition this interval into
three subintervals (0, $\Delta$), ($\Delta$, $R_\infty$) and
($R_\infty$, $\infty$), $\Delta$ being of the order of a few
atomic units  and $R_\infty$ being a point in the far asymptotic
domain. The solution in (0, $\Delta$) is then constructed as in
the R-matrix \cite{BR75} method and then continued to $R_\infty$
using Taylor's expansion method \cite{JND01,JND02}. Beyond
$R_\infty$ the solutions are known from series expansions
\cite{DPC03}. This approach, however, suffers from pseudo
resonance problems as pointed out in Ref. \cite{JND02} and hence
this is not the one adopted here. Other possibilities include
solving the set of equations as a two point boundary value problem
(since the radial wave function is known at origin, and at
$R_\infty$ from series expansions) as in the ECS method
\cite{BR01a}. This again would require more computational
resources than that we have at present. The most effective
approach for our purposes turns out to be the following. We
construct $N_{mx}$ independent solutions of Eq. (2) over the
interval (0, $\Delta$) by solving these as a two-point boundary
value problem. The k$^{th}$ solution vector is made to vanish at
the origin and takes the k$^{th}$ column of the $N_{mx}\times
N_{mx}$ unit matrix as its value at $\Delta$. These solutions are
then continued over ($\Delta$, $R_\infty$) by solving the coupled
system of equations by the Taylor's expansion method with frequent
stabilization \cite{CT75}. Beyond $R_\infty$ the solution may be
obtained from expansion in inverse powers of R with suitable sine
and cosine factors \cite{JND98,DPC03}. The asymptotic incoming
boundary conditions then completely define \cite{JND98,DPC03} the
scattering-state wave function $\Psi_{fs}^{(-)}$. For the initial
interval ( 0, $\Delta$ ) solution by the finite difference method
proves most effective. In our earlier calculation \cite{DPC03}, at
higher energies we used a five-point difference scheme. This gives
us very good cross sections for 30eV, 25eV, 19.6eV and 17.6eV for
various kinematic conditions. In our present calculation we
propose to use larger mesh size (double that of our previous
calculation) and hence, in place of a five-point difference scheme
we use a seven-point difference scheme. In the seven-point scheme
we divide the interval $[0, \Delta]$ into $m$ subintervals by
using mesh points $R_0, R_1, R_2, \cdots , R_{m-1}, R_m$ where
$R_k = hk, (k = 0, 1, 2, \cdots m)$ and $h =\Delta/m$. In this
scheme we use the following formulae:
\begin{widetext}
\begin{eqnarray}
f_N^{''}(R_k)&=& \frac{1}{\,h^2}[\ \frac{1}{90}f_N(R_{k-3})-
\frac{3}{20}16f_N(R_{k-2}) + \frac{3}{2}f_N(R_{k-1})-\frac{49}{18}
f_N(R_{k})+\frac{3}{2}f_N(R_{k+1}) -
\frac{3}{20}16f_N(R_{k+2})\nonumber\\
&+&\frac{1}{90}f_N(R_{k+3})] +
\{\frac{69}{25200}\,h^6f_N^{(viii)}(\xi_1)\}
\end{eqnarray}
\end{widetext}
for $k = 3, 4, \cdots , m-4, m-3$. For $k = 1, 2$ and $m-2, m-1$
we use the the formulae
\begin{widetext}
\begin{eqnarray}
f_N^{''}(R_1) &=&\frac{1}{h^2}[\ \frac{3}{8}f_N(R_0)+6f_N(R_1)-
\frac{11}{2}\,h^2f_N^{''}(R_2)-
\frac{51}{4}f_N(R_3)-\,h^2f_N^{''}(R_3)+6f_N(R_4)
+\frac{3}{8}f_N(R_4)\ ]\nonumber\\
&+& \{-\frac{23}{10080}\,h^6f^{(viii)}(\xi_2)\}.
\end{eqnarray}
\end{widetext}
\begin{widetext}
\begin{eqnarray}
f_N^{''}(R_2)&=&\frac{1}{h^2}[\ \frac{3}{8}f_N(R_1)+6f_N(R_2)-
\frac{11}{2}\,h^2f_N^{''}(R_3) -
\frac{51}{4}f_N(R_3)-\,h^2f_N^{''}(R_4)+6f_N(R_4)
+\frac{3}{8}f_N(R_5)\ ]\nonumber\\
&+& \{-\frac{23}{10080}\,h^6f^{(viii)}(\xi_3)\}.\\
f_N^{''}(R_{m-2}) &=&\frac{1}{h^2}[\
\frac{3}{8}f_N(R_{m-5})+6f_N(R_{m-4}) -\,h^2f_N^{''}(R_{m-4})
-\frac{51}{4}f_N(R_{m-3})-\frac{11}{2}\,h^2f_N^{''}(R_{m-3})\nonumber\\
&+&6f_N(R_{m-2})+ \frac{3}{8}f_N(R_{m-1})\ ] +
\{-\frac{23}{10080}\,h^6f^{(viii)}(\xi_4)\}\\
f_N^{''}(R_{m-1})&=&\frac{1}{h^2}[\
\frac{3}{8}f_N(R_{m-4})+6f_N(R_{m-3})-\,h^2f_N^{''}(R_{m-3})-
\frac{51}{4}f_N(R_{m-2})
-\frac{11}{2}\,h^2f_N^{''}(R_{m-2})\nonumber\\
&+&6f_N(R_{m-1})+\frac{3}{8}f_N(R_{m})\ ] +
\{-\frac{23}{10080}\,h^6f^{(viii)}(\xi_5)\}.
\end{eqnarray}
\end{widetext}
In each of Eqs. (3)-(7) quantities on the right hand sides within
the curly brackets represent the error terms. The corresponding
difference equations are obtained by substituting the values of
second order derivatives from the differential equation (2) into
these expressions. For continuing these solutions in the domain
$(\Delta, \; R_\infty)$ we need first order derivatives
${f'}_N(R)$ at $\Delta$. These are computed from the difference
formula

\begin{eqnarray}
f_N^{'}(R_m) & = & \frac{1}{84h}[-f_N(R_{m-4}) +
24f_N(R_{m-2})\nonumber\\
&-&128f_N(R_{m-1})+105f_N(R_m)]\nonumber\\
&+&\frac{2h}{7}f_N^{''}(R_m) + \{-\frac{4h^4}{105}f_N^{(v)}(\xi)\}
\end{eqnarray}

    Here too, the quantity within curly brackets represents the
error term. The solutions thus obtained in $(0, \Delta)$ are then
continued over $(\Delta, R_{\infty})$ by Taylor's expansion
method, as stated earlier, with stabilization after suitable steps
\cite{CT75}.

\section{Results}
In our present calculation for the equal-energy-sharing kinematics
and 1eV, 0.5eV and 0.3eV excess energies, we have included 30
channels and have chosen $\Delta$ = 5 a.u. (as in our previous
calculation \cite{DPC03} for higher energies). Small variation of
the value of $\Delta$ about the value chosen does not change the
results. Here we need to choose $R_\infty$ equal to 1000 a.u. for
1eV, 2000 a.u. for 0.5eV and 3000 a.u. for 0.3eV for smooth
convergence of the asymptotic series solution and for smooth
fitting with the asymptotic solution \cite{DPC03} in the equal
energy sharing cases. For unequal energy sharing cases one may
need to move to still larger distances. For going that far in the
asymptotic domain a larger value of $h$ (grid spacing) is
preferable. Here we have chosen $h = 0.1$ a.u. up to $\Delta$ and
a value 0.2 a.u. beyond $\Delta$ in all the cases. Accordingly a
seven point scheme, as described above, is more suitable over a
five point scheme used in our earlier calculation \cite{DPC03} and
hence we chose the above seven-point scheme in the present
calculation. We included states with L up to 5. Values of L above
5 give insignificant contributions. The ($l_1,l_2$) pairs which
have been included in our calculation are sufficient for
convergence as found from the results of calculations with the
inclusion of larger number of channels. All the computations were
carried out on a 2 CPU SUN Enterprise 450 system with 512 MB RAM.

For each incident energy three sets of results with different
geometry have been presented for the two outgoing electrons having
equal energies. These are the constant - $\theta_{ab}$ geometry
results, $\theta_{a}$-constant geometry results and the results
for symmetric geometry. For 1eV excess energy we have presented
these results in figures 1(a), 1(b) and 1(c) respectively. The
corresponding results for energy 0.5 eV are presented in figures
2(a), 2(b) and 2(c) respectively and the results for 0.3 eV have
been presented in figures 3(a), 3(b) and 3(c) respectively. The
bottom row of each of the figures 1(a), 2(a) and 3(a)
corresponding to $\theta_{ab}=150^o$ and $\theta_{ab}=180^o$ are
as in our earlier work \cite{DPC03a} (though these are now
calculated with different $R_\infty$ values). This is merely to
ensure completeness in the results presented and to compare our
results with the semiclassical calculation of Deb and Crothers.
For other geometries, unfortunately, neither experimental nor any
theoretical results are available for comparison. The overall
agreement between our results and those of Deb and Crothers
\cite{DC02} for $\theta_{ab} = 180^o$ and $\theta_{ab} = 150^o$ is
highly encouraging. A little steeper rise of our results compared
to those of Deb and Crothers \cite{DC02} at $0^o$ are in
conformity with the general trends of our corresponding earlier
results \cite{DPC03} at 2eV and 4eV excess energies. Our results
for $\theta_{ab} = 120^o$ and $100^o$ also appear reasonable when
compared with the shapes of the corresponding results for 2 eV and
4 eV excess energy \cite{DPC03}. Unfortunately there are no
experimental results for verification. The results for
$\theta_{a}$-constant geometry and those of symmetric geometry are
also in very nice agreement in shapes, particularly for
$\theta_{a}= -30^o$ and $\theta_{a} = -150^o$, with those for 2eV
and 4eV excess energy cases (see Das \textit{et al} \cite{DPC03}).
\section{Conclusion}
    From the results presented above it appears that the
hyperspherical partial wave theory works satisfactorily at 1eV,
0.5eV and 0.3eV excess energies. We have already good results
\cite{DPC03} for energies up to 30eV for various kinematic
conditions. Calculations at a higher incident energy of 54.4eV are
now in progress and we propose to present them in future. Another
point to note is that in this approach exploration of the far
asymptotic domain is possible by increasing $R_\infty$ to
thousands of atomic units. All these suggest that the
hyperspherical partial wave theory is capable of being developed
into a successful method for (e, 2e) collisions.
\begin{acknowledgments}
S.P. gratefully acknowledges a research fellowship provided by
CSIR.
\end{acknowledgments}

\bibliography{apssamp}

\begin{thebibliography}{50}
\bibitem{PC92}
C. Pan and A.F. Starace, Phys. Rev. Lett. 67, 185 (1991); Phys.
Rev. A 45, 4588 (1992).
\bibitem{JMS92}
S. Jones, D.H. Madison, and M. K. Srivastava, J. Phys. B
\textbf{25}, 1899 (1992).
\bibitem{RB99}
T. N. Rescigno, M. Baertschy, W. A. Isaacs, and  C. W. McCurdy,
Science \textbf{286}, 2474 (1999).
\bibitem{BR01a}
M. Baertschy, T. N. Rescigno, W. A. Isaacs, X. Li, and  C. W.
McCurdy, Phys. Rev. A \textbf{63}, 022712 (2001).
\bibitem{BR01b}
M. Baertschy, T. N. Rescigno,  and  C. W. McCurdy Phys. Rev. A
\textbf{64}, 022709 (2001).
\bibitem{RE97}
J. R\"oder, H. Ehrhardt, C. Pan, A. F. Starace, I. Bray, and D. V.
Fursa, Phys. Rev. Lett. \textbf{79}, 1666 (1997).
\bibitem{RB02}
J. R\"oder, M. Baertschy, and I. Bray, Phys. Rev A {\bf{67}},
010702(R) (2003).
\bibitem{RR96}
J. R\"oder, J. Rasch, K. Jung, C. T. Whelan, H. Ehrhardt, R. J.
Allan, and H. R. J. Walters, Phys. Rev. A \textbf{53}, 225 (1996).
\bibitem{BK94}
I. Bray, D. A. Konovalov, I. E. McCarthy, and A. T. Stelbovics
Phys. Rev. A\textbf{50}, R2818 (1994).
\bibitem{IB99}
I. Bray, J. Phys. B \textbf{32}, L119 (1999) ; 2000 J. Phys. B
\textbf{33}, 581 (2000).
\bibitem{IB00}
I. Bray, J. Phys. B \textbf{33}, 581 (2000).
\bibitem{DPC03}
J. N. Das, S. Paul, and K. Chakrabarti, Phys. Rev. A \textbf{67},
042717 (2003).
\bibitem{DC02}
N. C. Deb and D. S. F. Crothers, Phys. Rev. A \textbf{65}, 052721
(2002).
\bibitem{SM02}
P. Selles, L. Malegat, and A. K. Kazansky, Phys. Rev. A{\bf{65}},
032711 (2002).
\bibitem{JND98}
J. N. Das, Pramana- J. Phys. \textbf{50}, 53 (1998).
\bibitem{BR75}
P. G. Burke and W. D. Robb, Adv. At. Mol. Phys. \textbf{11}, 143
(1975).
\bibitem{JND01}
J. N. Das, Phys. Rev. A \textbf{64}, 054703 (2001).
\bibitem{JND02}
J. N. Das, J. Phys. B  \textbf{35}, 1165 (2002).
\bibitem{LIN74}
C. D. Lin, Phys. Rev. A \textbf{10}, 1986 (1974).
\bibitem{CT75}
B. H. Choi and K. T. Tang, J. Chem. Phys. \textbf{63}, 1775
(1975).
\bibitem{DPC03a}
J. N. Das, S. Paul, and K. Chakrabarti, AIP Conference Proceedings
\textbf{697}, 82 (2003).
\end{thebibliography}

\begin{widetext}
\underline{\bf{Figure Captions}}\\

\noindent \textbf{Figure 1(a).} TDCS for coplanar
equal-energy-sharing constant $\Theta_{ab}$ geometry at 1eV excess
energy above threshold. Continuous curves, present results ;
dashed-curves, semiclassical results of Deb and Crothers \cite{DC02}\\

\noindent \textbf{Figure 1(b).} TDCS for coplanar
equal-energy-sharing geometry at 1eV excess energy above threshold
for fixed $\theta_a$ and variable $\theta_b$ of the out going electrons.\\

\noindent \textbf{Figure 1(c).} TDCS for coplanar
equal-energy-sharing with two electrons emerging on opposite sides
of the direction of the incident electron with equal angle $\theta_a$
at 1eV excess energy above threshold.\\

\noindent \textbf{Figure 2(a).} TDCS for coplanar
equal-energy-sharing constant $\Theta_{ab}$ geometry at 0.5eV
excess energy above threshold. Continuous curves, present results
; dashed-curves, semiclassical results of Deb and Crothers \cite{DC02}\\

\noindent \textbf{Figure 2(b).} TDCS for coplanar
equal-energy-sharing geometry at 0.5eV excess energy above
threshold for fixed $\theta_a$ and variable $\theta_b$ of the out
going electrons.\\

\noindent \textbf{Figure 2(c).} TDCS for coplanar
equal-energy-sharing with two electrons emerging on opposite sides
of the direction of the incident electron with equal angle
$\theta_a$ at 0.5eV excess energy above threshold.\\

\noindent \textbf{Figure 3(a).} TDCS for coplanar
equal-energy-sharing constant $\Theta_{ab}$ geometry at 0.3eV
excess energy above threshold. Continuous curves, present results
; dashed-curves, semiclassical results of Deb and Crothers \cite{DC02}\\

\noindent \textbf{Figure 3(b).} TDCS for coplanar
equal-energy-sharing geometry at 0.3eV excess energy above
threshold for fixed $\theta_a$ and variable $\theta_b$ of the out
going electrons.\\

\noindent \textbf{Figure 3(c).} TDCS for coplanar
equal-energy-sharing with two electrons emerging on opposite sides
of the direction of the incident electron with equal angle
$\theta_a$ at 0.3eV excess energy above threshold.\\

\end{widetext}

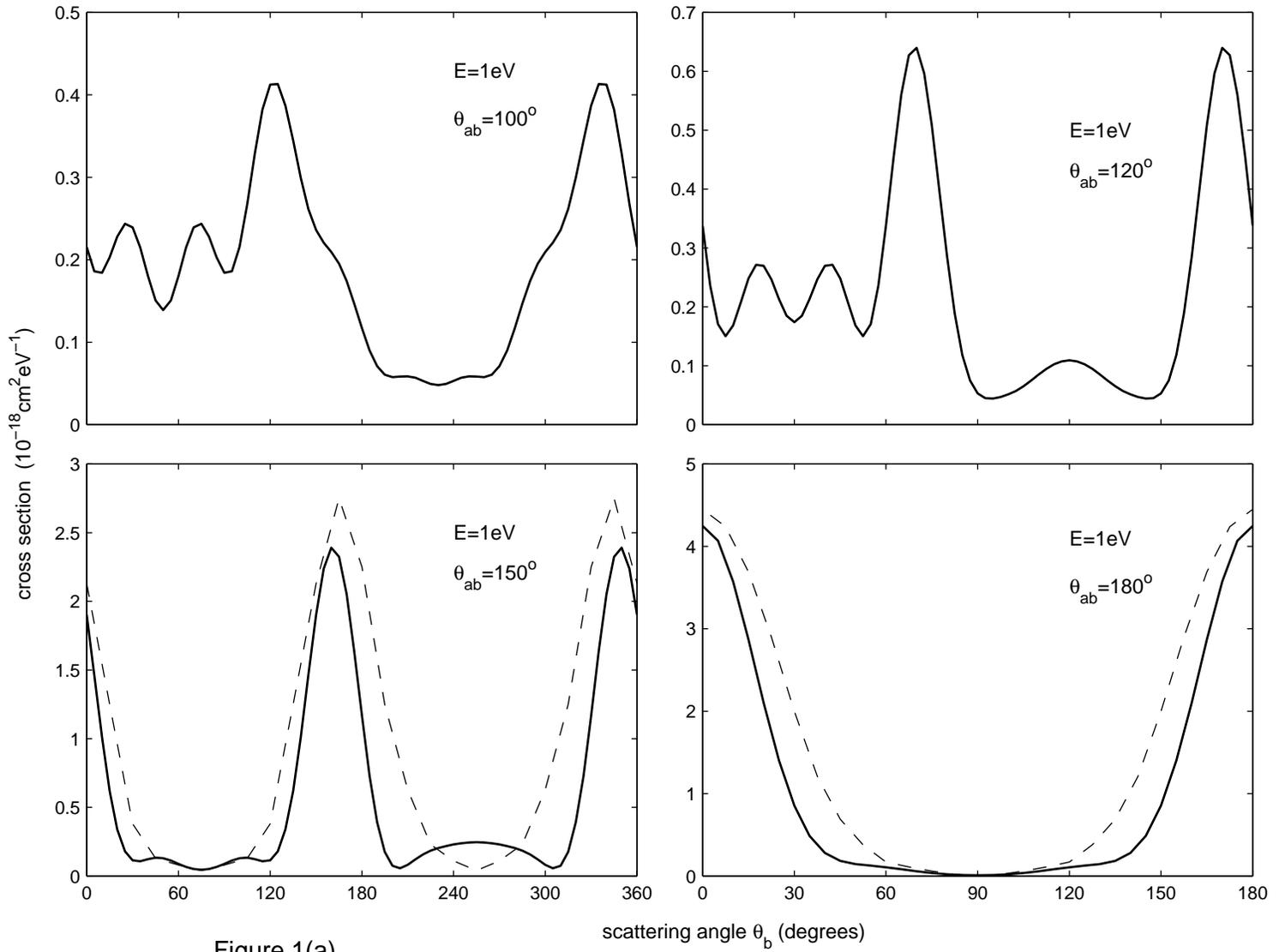
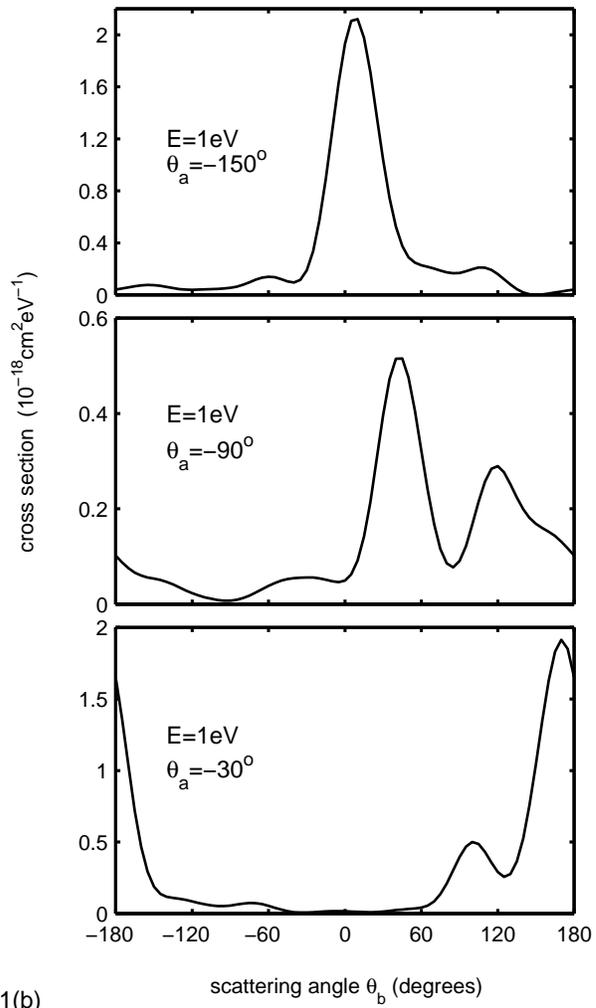
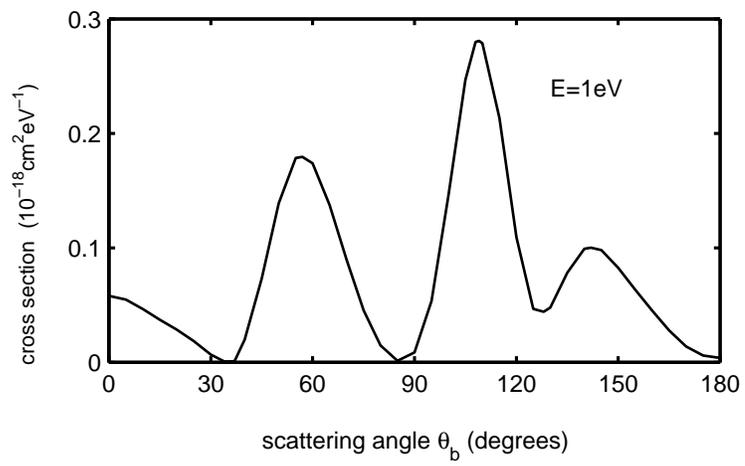
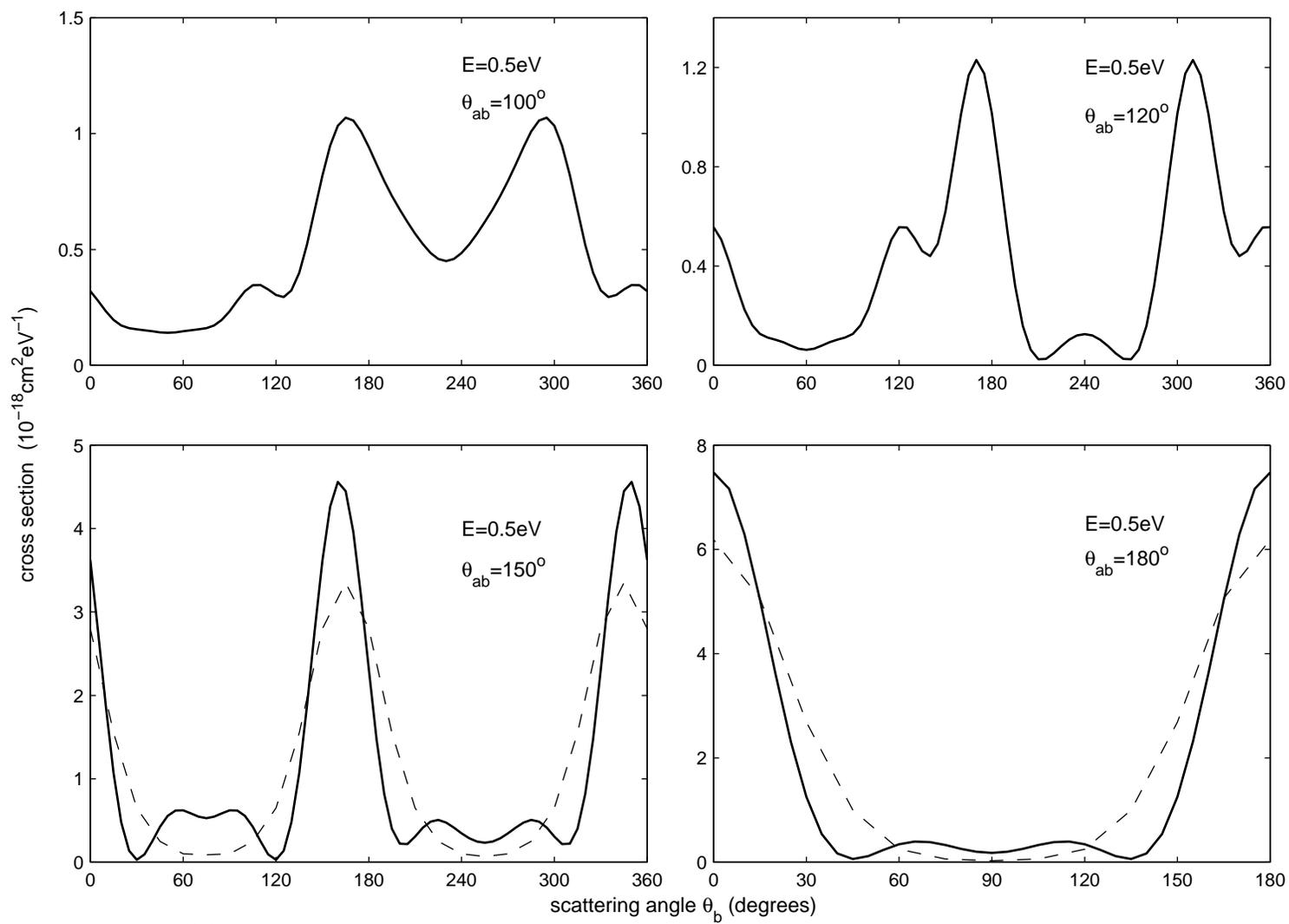
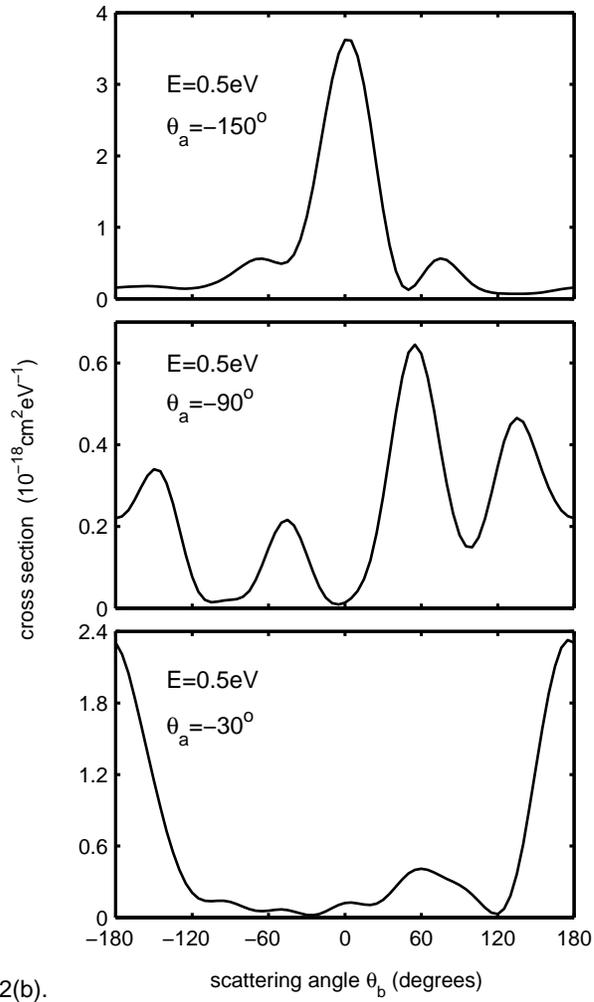
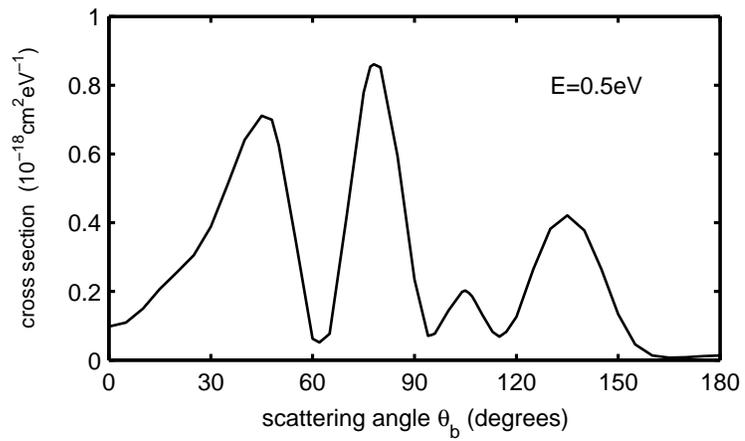
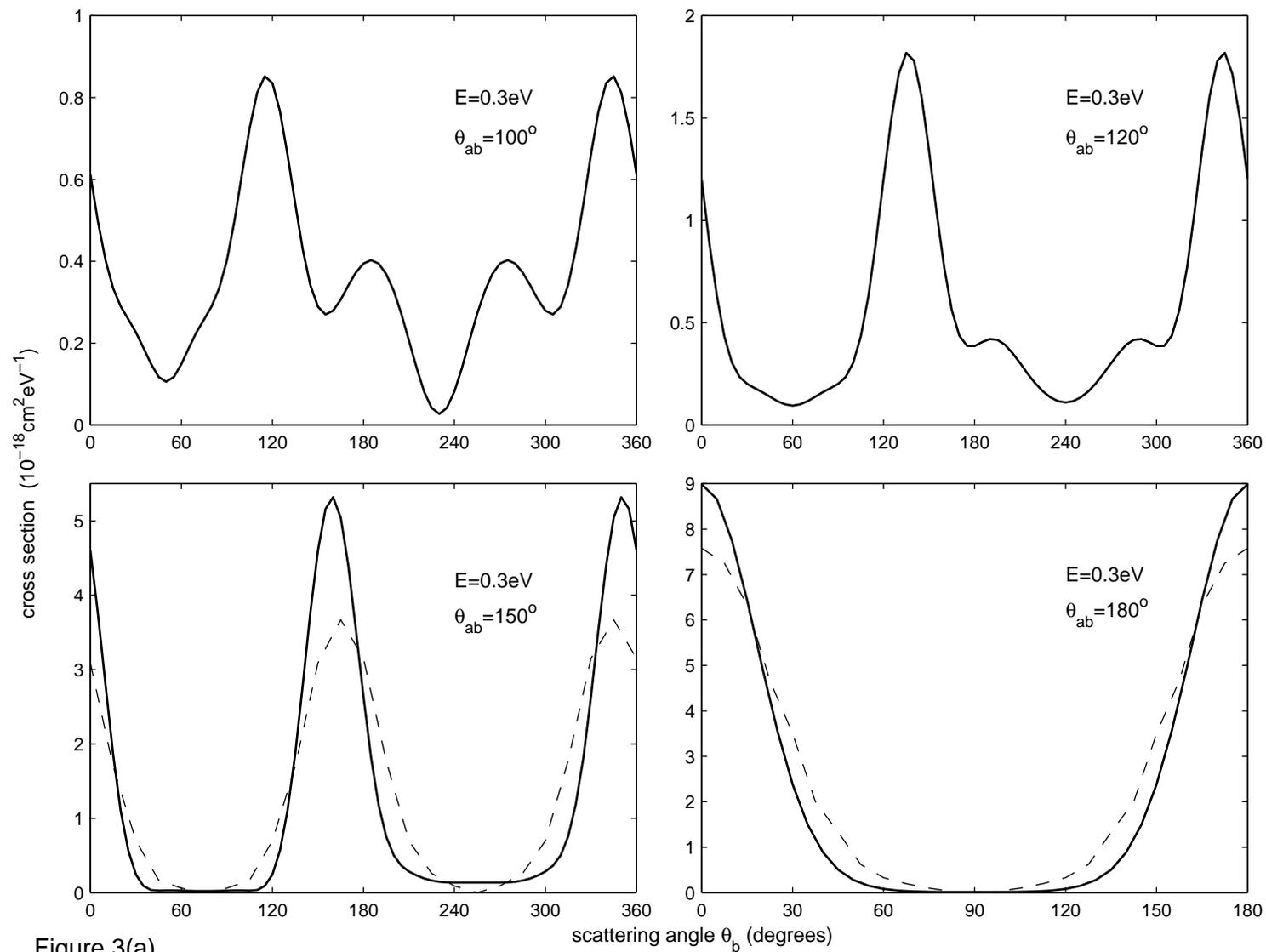
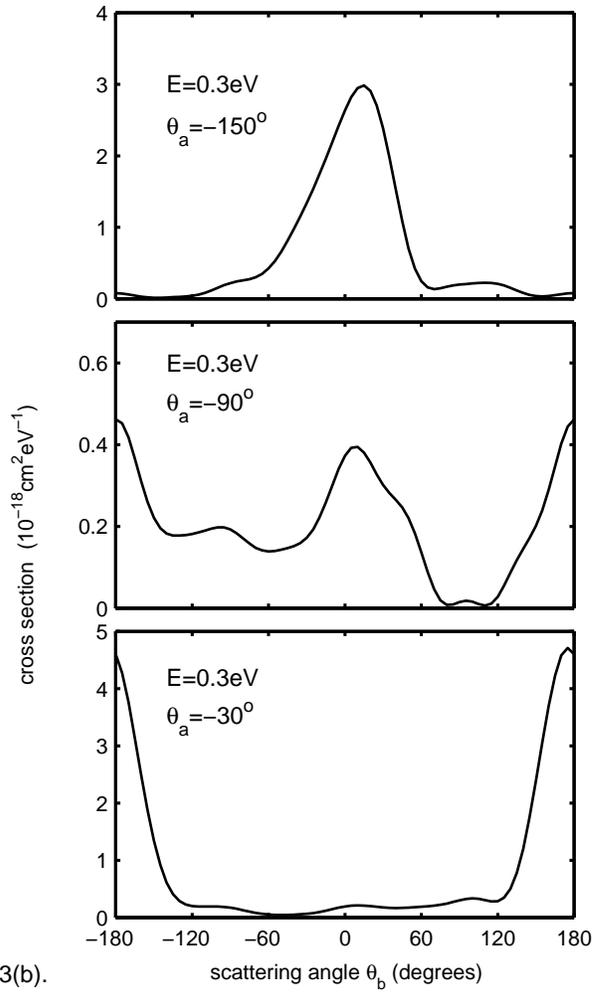
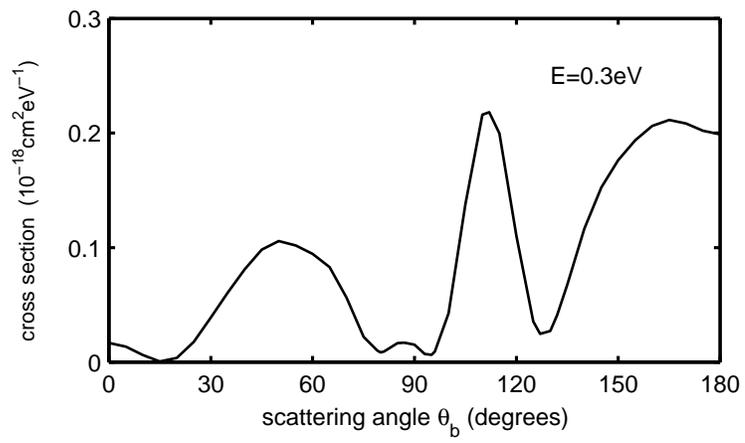
\end{document}